\documentclass[journal=ancac3,layout=twocolumn]{achemso}
\usepackage{hyperref}
\usepackage{amssymb}
\usepackage{amsmath}
\usepackage{graphicx}

\newcommand{\onlinecite}[1]{\hspace{-1 ex} \nocite{#1}\citenum{#1}}

\title{\large Nanograined half-Heusler semiconductors as advanced thermoelectrics: \\ an {\itshape ab-initio} high-throughput statistical study}

\author{J. Carrete}
\author{N. Mingo}
\email{natalio.mingo@cea.fr}
\affiliation{CEA-Grenoble, 17 Rue des Martyrs, Grenoble 38000, France}

\author{S. Wang}
\author{S. Curtarolo}
\email{stefano@duke.edu}
\affiliation{Center for Materials Genomics, Materials Science, Electrical Engineering, Physics and Chemistry, Duke University, Durham, North Carolina 27708, USA}

\begin{document}

\begin{abstract}
  Nanostructuring has spurred a revival in the field of direct thermoelectric energy conversion.
  Nanograined materials can now be synthesized with higher figures of merit ($ZT$) than the bulk counterparts.
  This leads to increased conversion efficiencies.
  Despite considerable effort in optimizing the {\it known} and discovering the {\it unknown}, technology still relies upon a few limited solutions.
  Here we perform {\it ab-initio} modeling of $ZT$ for $75$ nanograined compounds obtained by filtering down the 79,057 half-Heusler entries available in the {\small \sf AFLOWLIB}.org 
  repository according to electronic and thermodynamic criteria.
  For many of the compounds the $ZT$s are markedly above those attainable with nanograined IV and III-V semiconductors.
  About $15\%$ of them may even outperform $ZT\sim 2$ at high temperatures.
  Our analysis elucidates the origin of the advantageous thermoelectric properties found within this broad material class.
  We use machine learning techniques to unveil simple rules determining
  if a nanograined half-Heusler compound is likely to be a good thermoelectric
  given its chemical composition.
\end{abstract}

\section*{Introduction}

Harnessing the thermoelectric effect to scavenge electric power from waste heat has long been an attractive route in the pursuit of
sustainable energy generation \cite{snyder_complex_2008}.
Despite recent progress, the goal of producing efficient thermoelectric materials remains elusive due to several challenging
factors \cite{snyder_complex_2008,curtarolo:nmat_review}.
Effective thermoelectrics must have a high thermoelectric figure of merit \cite{snyder_thermoelectric_2003,Vining_NM_2009}:
\vspace{-1.5mm}
\begin{equation}
  ZT=\frac{\sigma S^2 T}{\kappa},
  \label{eqn1}
\end{equation}
\noindent where $T$, $\sigma$, $S$ and $\kappa$ are temperature, material's electrical conductivity, Seebeck coefficient and thermal conductivity, respectively.
$\kappa$, in turn, can typically be split into a sum of vibrational ($\kappa_{\ell}$) and electronic ($\kappa_{\mathrm{e}}$) contributions.
 To design useful thermoelectric materials, the power factor ($P=\sigma S^2$) must be improved and the thermal conductivity reduced.

Nanostructuring approaches may significantly improve $ZT$ \cite{li_high-performance_2010}. Still, significant practical roadblocks remain to their production.
Low-dimensional nanostructures (e.g. nanowires, thin films) methods can only be used in mesoscopic and macroscopic devices, and only after
an assembly process demanding additional engineering and/or packaging \cite{curtin_highly_2012}.
Embedding nanophases to enhance bulk thermoelectric properties poses
challenges due to controlling the size and morphology of precipitates \cite{pei_high_2011}.
A third alternative is the use of {\itshape ad-hoc} nanocrystalline bulk materials having nanoscale grains \cite{bux_nanostructured_2009,lan_structure_2009},
core-shell structures \cite{Poon_coreshell_APL_2013}, or undergoing spinodal decomposition \cite{Sootsman_JAP_2009,Gelbstein_CHEMMAT_Spinodal_2010}. 
Such intrinsic features might ease synthesis.

Due to the high cost of experiments, studies of nanostructured materials for thermoelectric applications tend to focus on known compounds
whose bulk properties are already promising.
Well-known examples are Bi$_2$Te$_3$ and PbTe \cite{Poudel2008,biswas_high-performance_2012}. 
Alternatively, some researchers choose inexpensive and widely available materials, such as Si, and try to optimize
their performance \cite{hochbaum_enhanced_2008}.
Very few studies, if any, have attempted to examineradically new materials from scratch \cite{madsen2006}.

\begin{figure}[t!]
  \begin{center}
    \includegraphics[width=\columnwidth]{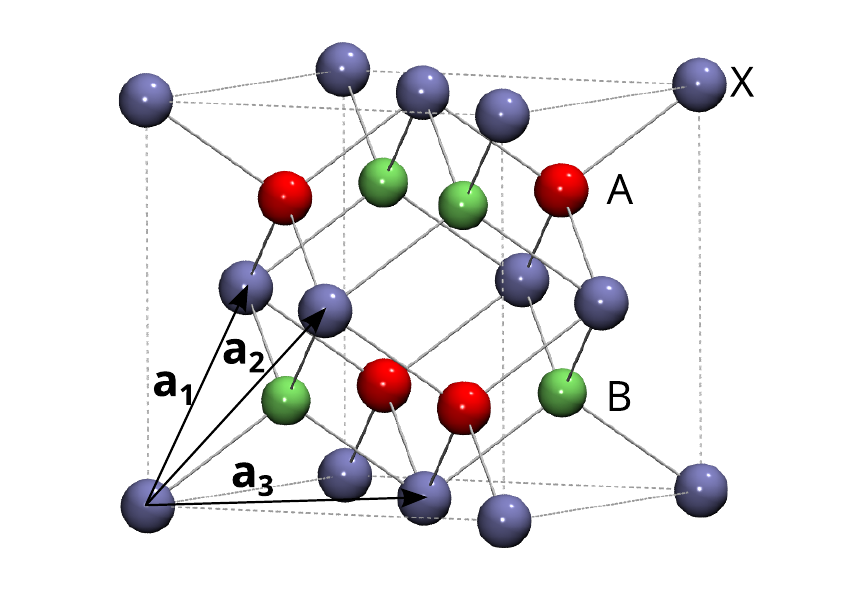}
  \end{center}
  \vspace{-10mm}
  \caption{\small The cubic conventional cell of the half-Heusler structure (cF12).
    The three primitive vectors, $\{\mathbf{ a_1},\mathbf{ a_2},\mathbf{ a_3}\}$, identify the standard primitive cell \cite{aflowBZ}.}
  \label{fig1}
\end{figure}

This work presents the first fully {\itshape ab-initio} exploration of $ZT$ for a large library of materials in a nanostructured configuration.
We focus on half-Heusler (HH) compounds due to
their typically high bulk power factors \cite{snyder_thermoelectric_2003}, and the fact that many possible compositions are still unexplored.
HH compounds are ternary $XAB$ solids.
 Their crystalline structure consists of three inter-penetrated fcc lattices.
Fig. \ref{fig1} depicts the conventional and primitive cells of such structures.

HHs are Heusler systems with a vacancy in one of the two doubly degenerate sublattices.
 This advantageous vacancy allows HHs
to be easily doped, and their properties manipulated.
Solubility limitations from size/electronegativity/character seen in other crystallographic prototypes are avoided \cite{curtarolo:art41}.
Extensive studies on bulk forms or with nanoinclusions have been performed on a tiny number of HH alloys \cite{xie_recent_2012,Yan_NL_2011,poon_half-heusler_2011} 
(e. g. NiSn$_x$Bi$_{1-x}$Zr$_y$Hf$_{1-y}$  Ref. \onlinecite{uher_thermoelectric_1999} and CoSn$_x$Sb$_{1-x}$Zr$_y$Hf$_{1-y}$ Ref. \onlinecite{culp_zrhfcosbsn_2008}).

Theoretical calculations have yet to characterize these compounds to their fullest extent \cite{yang_evaluation_2008,Wee_Fornari_TiNiSn_JEM_2012}.
Neighboring fields of research (e.g. spintronics, magnetism, topological insulators) are actively investigating
Heusler systems via experimental and high-throughput approaches
(Refs. \onlinecite{matsumoto2003combinatorial,curtarolo:nmat_review,curtarolo:TIs} and references therein).

We start by considering the $79,057$ HH compounds included in the {\small \sf AFLOWLIB}.org consortium repository \cite{aflowlibPAPER,aflowPAPER}. 
This list contains all conceivable unique half-Heusler structures that can be built using elements from the list 
\{Ag, Al, As, Au, B, Ba, Be, Bi, Br, Ca, Cd, Cl, Co, Cr, Cu, Fe, Ga, Ge, Hf, Hg, In, Ir, K, La, Li, Mg, Mn, Mo, Na, Nb, Ni, Os, P, Pb, Pd, Pt, Re, Rh, Ru, Sb, Sc, Se, Si, Sn, Sr, Ta, Tc, Te, Ti, Tl, V, W, Y, Zn, and  Zr\}. 
From this set, only compounds whose formation enthalpies are negative and optimal with respect to elements' permutations are kept. 
We then discard any metallic compounds.

Supercell calculations are run to obtain full phonon dispersion relations of the remaining half-Heuslers.
Only $450$ compounds with fully real dispersions over the whole Brillouin zone (BZ) are kept.
Mechanically unstable configurations are thus neglected.

Finally, the ternary phase diagrams for each of the $450$ mechanically stable HHs are obtained from {\small \sf AFLOWLIB}.org to assess their thermodynamical stability.
More than $110,000$ elemental, binary and ternary phases are taken into account in these diagrams.
Included are all relevant binary and ternary phases from the Inorganic Crystal Structure Database (ICSD) \cite{ICSD,ICSD4}.
Only the $77$ thermodynamically stable HHs remain after these selection criteria are applied. 
Spin-polarized calculations reveal that two of the $77$ have semimetallic ground states. Only the remaining $75$ are chosen for further study. Note that checking our results against all the possible competing ground states from the ICSD database does not prevent other non-HH phases
to appear at higher temperature, through vibrational-, magnetic- or configurational-entropic stabilization (e.g., NiMn$_{1−t}$Ti$_t$Ge, Ref. \onlinecite{NiMnTiGe_Bazela_1981}).
The ICSD may also contain inaccuracies and is not guaranteed to be comprehensive. Ground states not listed in the ICSD can be found by global optimization methods poorly suited for high-throughput applications. However, in some cases our calculations predict the half-Heusler to be more stable than the non-HH isostoichometric phase reported in the ICSD. Specifically, this happens for CoAsHf, CoGeNb, CoGeTa, CoGeV, CoNbSi, CoSiTa, FeAsNb, FeAsTa, IrGeNb, NiGeHf, NiGeTi, NiGeZr, PdGeZr, PtGeZr, PtLaSb, RhAsTi and ZnLiSb. The ternary phase diagram of each of the 75 ternary compounds studied here can be found in the supplemental material to Ref. \onlinecite{curtarolo:art84}. 

The high density of grain boundaries in nanograined materials has a strong influence in the physics of thermal transport therein.
Phonon and electron scattering at grain boundaries adds a contribution to the scattering probability density. 
This contribution is inversely proportional to the characteristic grain size \cite{klemens_phonon_1994,callaway_model_1959,aflowTHERMO}. In the nanograined limit, this contribution dominates 
over the factors that typically determine thermal conductivity in bulk semiconductors and alloys: three-phonon processes \cite{mingo_ab_2014}, isotopic disorder \cite{tamura_isotope_1983}, alloying and embedded particles \cite{kundu_role_2011}.
All mean free paths can be approximated by a single value $\lambda$, which is the same order of magnitude as the grain size. 
The general expressions for the phenomenological transport coefficients \cite{ziman_electrons_2001} can then be reduced to an approximate simplified form (see Methods section). 
This allows $\sigma$, $\kappa_{\mathrm{e}}$ and $\kappa_{\ell}$ to become proportional to $\lambda$, while $S$ remains independent. After manipulation of Eq. (1), $ZT$ reduces to a form independent from $\lambda$.

Hence, full knowledge of electronic and vibrational band structures for each compound becomes the essential
ingredient for computing $ZT$.
This approach needs no further approximations.
The results are expected to be robust compared to transport calculations in other regimes.

An important issue is whether the nanograined limit can be achieved with realistic grain sizes.
We estimate $\lambda$ by comparing our calculated coefficients with the bulk values.
For electrons, even in alloys, published mobilities are linked to mean free paths in the $10-100\,\mathrm{nm}$ range \cite{xie_recent_2012}.
For phonons, we calculate the bulk thermal conductivity of several tens of HHs using the Boltzmann transport equation.
We find that typically $\lambda\!\gtrsim\!10\,\mathrm{nm}$ \cite{aflowKAPPA} at room temperature, and that it only decreases slightly at high temperatures.
Thus, in parallel to what has been proven for Si \cite{bux_nanostructured_2009}, the small-grain-size limit can be experimentally approached with nanograined HHs.
This may allow for the development of high-performance thermoelectrics.

This approach to the nanograined-limit is a convenient simplification. A realistic description of grain boundaries would have to take into account the geometrical, chemical and electrical structures of those interfaces, which can give rise to energy- and charge-dependent scattering (see Refs. \onlinecite{wang_thermal_2011,bachmann_ineffectiveness_2012} and references therein). Those effects are likely to depend on the sample preparation procedure. On the other hand, having a simple descriptor like the nanograined-limit $ZT$ is optimal for comparing compounds from a high-throughput perspective with respect to a single factor --- nanograining. The use of a constant mean free path to describe boundary scattering has a long tradition\cite{parrott_thermal_1969}; more recent, finer studies support the idea that a best-fit constant mean free path can be found, although is likely to underestimate the experimental grain size \cite{wang_thermal_2011}.

\section*{Results and Discussion}
\begin{figure}[t!]
  \begin{center}
    \includegraphics[width=\columnwidth]{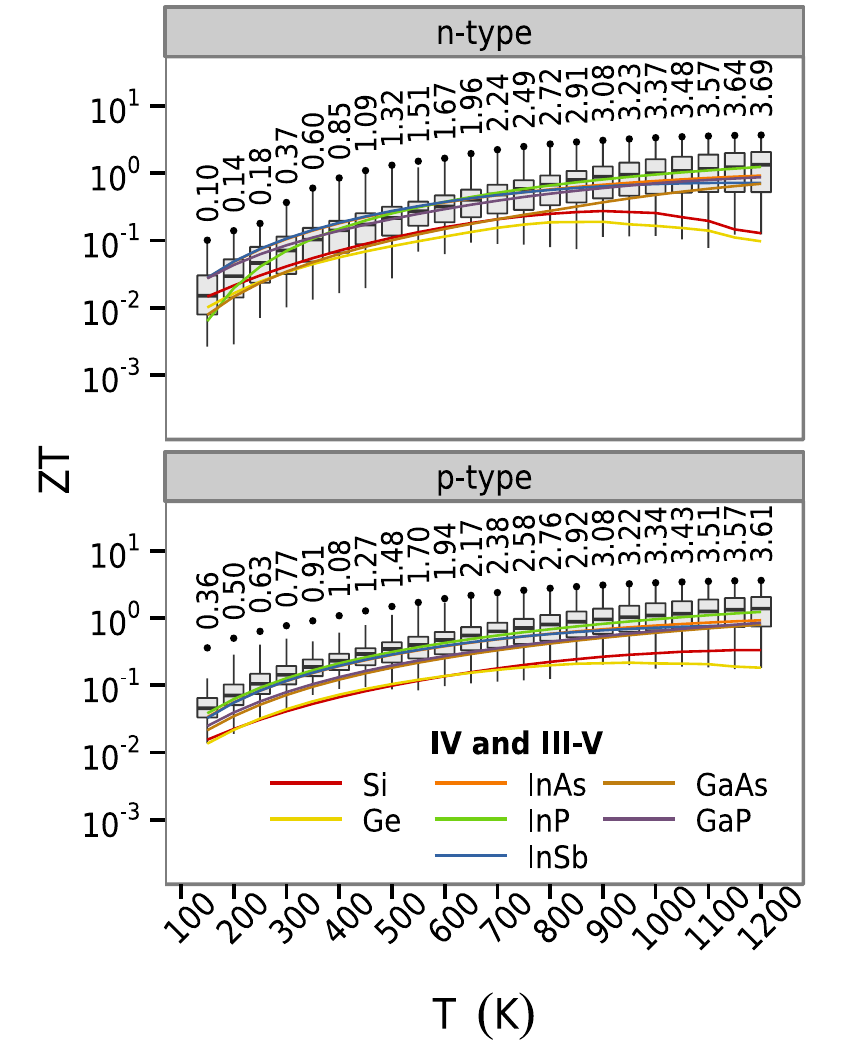}
  \end{center}
  \vspace{-5mm}
  \caption{\small
    Distributions of nanograined $ZT$ values for the HH compounds, at different temperatures. Boxes extend from the first to the third quartile,
    the median is represented by a horizontal black segment and whiskers have a length equal to $1.5$ times the interquartile range.
    The maximum $ZT$ at each temperature is drawn as a circle and its value annotated on the plot.
 Data for some elemental group-IV and binary III-V semiconductors are plotted for reference.}
  \label{fig2}
\end{figure}

The temperature distributions of nanograined $ZT$s are presented in Fig. \ref{fig2}, for both the $n$-type (top panel) and $p$-type (bottom panel) doping regimes.
 The maximum carrier concentration allowed in the calculation is $10^{21}\,\mathrm{cm^{-3}}$ to remain within the doping limits typical of experimental
HH thermoelectrics \cite{snyder_complex_2008} (Methods section).
Detailed values for all compounds are included in the supporting information.
 For comparison, we also perform calculations for Si, Ge and five common III-V binary semiconductors,
InAs, InP, InSb, GaAs and GaP; their results are superimposed in both panels.
An experimental study \cite{bux_nanostructured_2009} on the properties of nanostructured bulk Si reports values compatible with our estimation: for instance, at $500\,\mathrm{K}$
the reported $ZT$ varies between $0.065$ and $0.087$ depending on milling conditions, which corroborates our $0.075$ result. 

The best considered elemental and binary thermoelectric semiconductors
are InP and InSb, yet their corresponding curves in Fig. \ref{fig2} are usually around the median and never surpass the third quartile of the HH distribution.
This indicates that most of the HHs outperform them at any given temperature.
 Indeed, many HH improve on InP by a very significant margin.
The highest values of $ZT$ are competitive with or better than the best reported for nanostructured bulk systems \cite{li_high-performance_2010}.
Especially remarkable are the values in excess of {$2$} achieved for $T>600\,\mathrm{K}$.

Comparison of the panels in Fig. \ref{fig2} reveals that a typical ({\itshape i.e.} close to the median) $p$-type-doped HH is, at all but the highest temperatures,
a much better thermoelectric than a typical $n$-type-doped HH.
 In contrast, the best $n$-type doped
 compounds are comparable with the best among the $p$-type-doped.
 Notably, the fraction of compounds achieving optimal thermoelectric efficiencies when $n$-doped
 increases monotonically with temperature.
 This ranges from just {$13\%$} at $300\,\mathrm{K}$, to $20\%$ at $1000\,\mathrm{K}$.

Qualitatively, these phenomena can be explained
using a two-parabolic-band model.
According to our band structure calculations, for $65\%$ of compounds the effective mass of holes ($m_h^*$) is higher than that
 of electrons ($m_e^*$).
 This implies a generally higher optimized power factor for the $p$-type \cite{aflowTHERMO}.
 This explains the general trend of the medians, and allows enough exceptions to account for the maximum values.

\begin{figure}[t!]
  \begin{center}
    \includegraphics[width=\columnwidth]{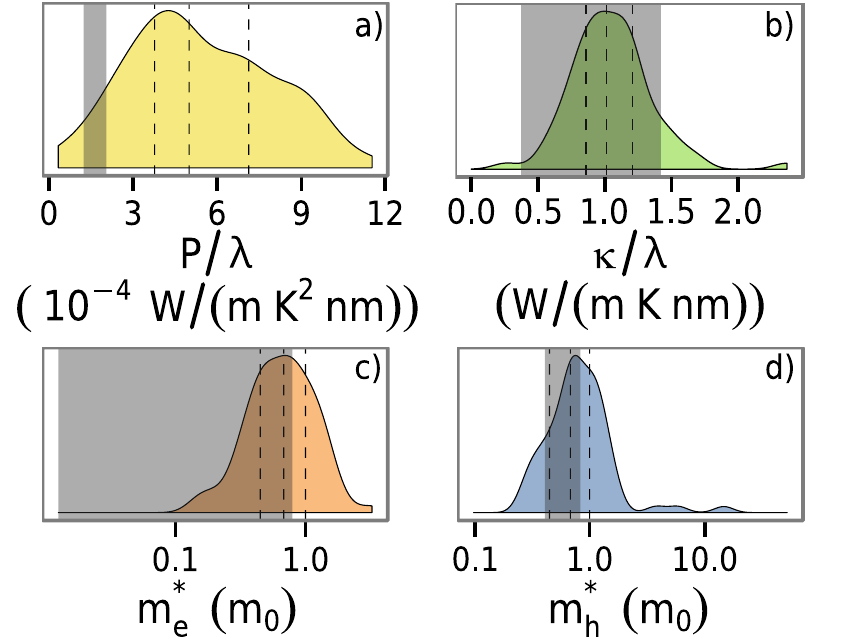}
  \end{center}
  \vspace{-5mm}
  \caption{\small
    Frequency distributions among the {$75$} HHs of:
    {\bfseries a)} power factors divided by grain size at $300\,\mathrm{K}$,
    {\bfseries b)} total thermal conductivities divided by grain size at $300\,\mathrm{K}$,
    {\bfseries c)} effective masses of electrons and {\bfseries d)} effective masses of holes.
    Dashed vertical lines show the positions of the quartiles.
 The shaded areas span the values of each variable among the other seven semiconductors listed in Fig. \ref{fig2}.}
  \label{fig3}
\end{figure}

Here we choose the best possible doping and try to reveal the factors contributing to improved performance.
The power factor  is key for selecting out these HHs from all other  semiconductors considered, as panels (a) and (b) in Fig. \ref{fig3} show.
While the thermal conductivities of the nanograined IV and III-V compounds are comparable to the HHs,
their power factors barely reach the tenth percentile (most likely due to the fact that effective masses
can be much higher in HHs).

We choose $300\,\mathrm{K}$ and $1000\,\mathrm{K}$ as representative of room- and high-temperature behaviors.
We unfold some of the factors that determine desirable HH thermoelectrics.
For this we use the Spearman rank correlation coefficient, $\Sigma$, which serves as an indicator of monotonic relationships between variables.
We compute $\Sigma$ among all the thermoelectric coefficients entering Eq. (1).
We also compute $\Sigma$ between them and a data
set of compound descriptors.
 These descriptors include the bandgap $\epsilon_g$,
the lattice parameter $a_{\mathrm{0}}$ and the aforementioned effective masses.

We find that at $300\,\mathrm{K}$, $ZT$ and $\kappa_e$ have a $\Sigma$ value of {$-0.08$}.
This becomes $-0.35$ at $1000\,\mathrm{K}$.
The increase in absolute value arises from the bigger role of heat transfer by charge carriers at high temperatures.
The median contribution of charge carriers to $\kappa$ increases from {$5\%$} at $300\,\mathrm{K}$ to $33\%$ at $1000\,\mathrm{K}$.
Unsurprisingly, $P/\lambda$ is a better predictor of $ZT$ than $\kappa$.
This is due to its wider range of values ($\Sigma_{300\,\mathrm{K}}=0.86$ and $\Sigma_{1000\,\mathrm{K}}=0.89$ between $P/\lambda$ and $ZT$;
  \textit{versus} $\Sigma_{300\,\mathrm{K}}=-0.25$ and $\Sigma_{1000\,\mathrm{K}}=-0.44$ between $\kappa/\lambda$ and $ZT$).
  
At both temperatures, the variation of the lattice contribution to $\kappa$ over the set of compounds may be explained by the monotonically decreasing dependence
of $\kappa_{\ell}$ on $a_{\mathrm{0}}$ ($\Sigma_{300\,\mathrm{K}}=-0.79$ and $\Sigma_{1000\,\mathrm{K}}=-0.81$).
This is understandable taking the equipartition theorem into account,
and considering the similitude of the expression for $\kappa_{\ell}$ in Eq. (2d) with that of the vibrational specific heat.
Likewise, the power factor exhibits a similar behavior at both room and high temperatures: $P/\lambda$ depends most markedly on 
{$m_h^*$ ($\Sigma_{300\,\mathrm{K}}=0.85$ and $\Sigma_{1000\,\mathrm{K}}=0.79$) and on} $\epsilon_g$ ({$\Sigma_{300\,\mathrm{K}}=\Sigma_{1000\,\mathrm{K}}=0.70$}). 
This reflects the advantage of {maximizing the contribution of majority carriers while} keeping the minority carriers as unexcited as possible.

{\bfseries Simple recipes for HH thermoelectrics}.
Simple recipes are key factors for the transfer of theoretical results into practical technology.
Rather than relying on intuition, we opt for the use of machine learning techniques. These objective methods are capable of discovering potentially hidden rules.

\begin{figure}[t!]
  \begin{center}
    \includegraphics[width=\columnwidth]{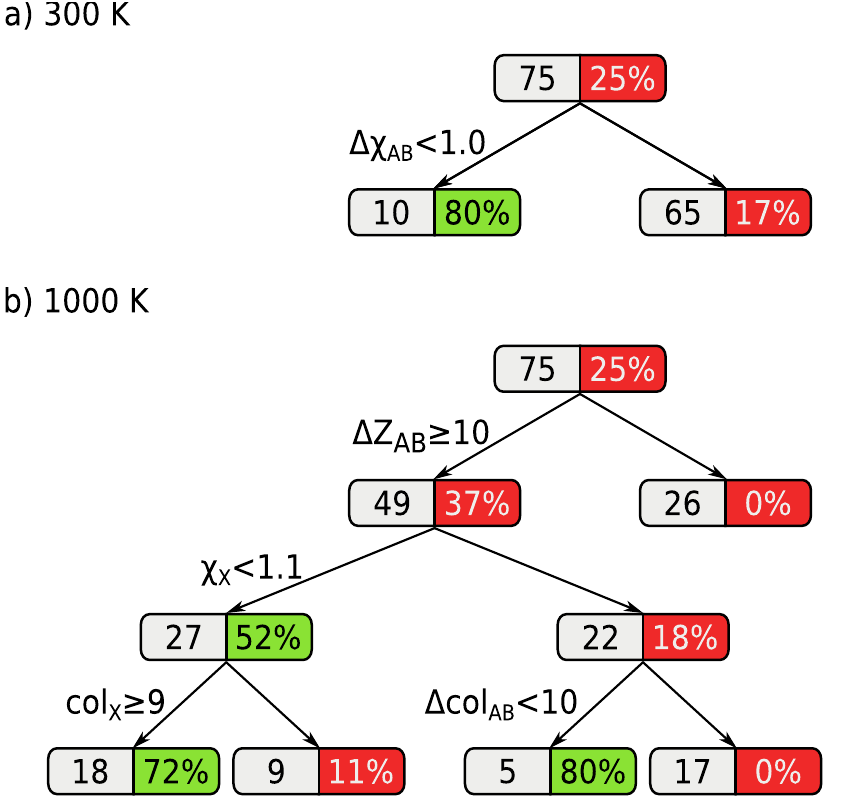}
  \end{center}
  \vspace{-5mm}
  \caption{\small
    Classification trees for HH compounds at $300\,\mathrm{K}$ (top) and $1000\,\mathrm{K}$.
    Printed on each node are the number of compounds it contains, and the fraction of them above the third quartile of the $ZT$ distribution.
    {Left-leaning} branches correspond to increased percentages of compounds in the best-$ZT$ category.}
  \label{fig4}
\end{figure}

First, we define the best thermoelectrics at a given temperature as those whose $ZT$ is beyond the third quartile of the distribution.
These are either {$ZT>0.20$} at $300\,\mathrm{K}$, or {$ZT>1.66$} at $1000\,\mathrm{K}$.
Second, we then consider a few elements' properties $\left\{\psi^i\right\}$:
atomic numbers ($Z$) and masses, positions in the periodic table (colum/row, $col/row$),
atomic radii ($r$), Pauling electronegativities \cite{pauling_bond}
and Pettifor's chemical scales ($\chi$) \cite{pettifor:1984}.
Third, from each property $\psi$ we build three descriptors:
the property value for element $X$, $\psi^i_X$;
the averaged property for elements $A,B$, $\bar{\psi}^i_{AB}\equiv\left(\psi^i_A+\psi^i_B\right)/2$ (preserving choice $A\leftrightarrow B$);
the absolute property difference for elements $A,B$, $\Delta \psi^i_{AB} \equiv \left\vert \psi^i_A-\psi^i_B\right\vert$.
Fourth, we grow a decision tree classifying the compounds in the best-$ZT$ category by performing a binary split of the data minimizing the Gini impurity increase \cite{breiman_classification_1984}.
As a stop-branching criterion, {no further splits are attempted within nodes with less than $15$ compounds}.
Finally, the fully grown tree is pruned back to the point where the cross-validation error is minimized. 
The procedure converges to the trees of Fig. \ref{fig4}. 
The room- and high-temperature trees put $83\%$ and $89\%$ of the compounds in the correct group, respectively.

The key variable in each classification is the one at the top of each tree {Although high thermoelectric efficiencies at room and at high temperature are highly correlated ($\Sigma=0.82$ between $ZT_{300\,\mathrm{K}}$ and $ZT_{1000\,\mathrm{K}}$) this difference shows that the most efficient strategy to sample only the top of the distribution depends on $T$. At $300\,\mathrm{K}$ the best criterion is to choose two elements with similar values in Pettifor's chemical scale as $A$ and $B$.} Through considerations of electron-phonon scattering, a similar rule ---a negative correlation between differences in electronegativity and good bulk thermoelectric properties--- has been suggested by several authors \cite{slack_crc_1995,Mahan_review_TE}. Here the situation is different. In nanograined materials the dominant scattering mechanism is, by construction, grain boundary scattering. 
Thus, only electronic and phononic band structures matter. In this case, an analysis of the percentiles of $P/\lambda$ and $\kappa/\lambda$ of compounds in in the best-$ZT$ category with respect to the percentiles of the full population shows that the best nanograined thermoelectrics at $300\,\mathrm{K}$ outperform the rest mainly because of their lower lattice thermal conductivity. This can be attributed to their larger lattice constants, as explained above. A possible explanation of the connection between $\Delta\chi_{AB}$ and $a_{\mathrm{0}}$ is that compounds with a low $\Delta\chi_{AB}$ tend to be formed by elements from the regions of the periodic table with the largest atomic radii.

At $1000\,\mathrm{K}$, however, it is advantageous to select out HHs where $A$'s and $B$'s atomic numbers differ by less than $10$. 
  Alone, the condition is not sufficient to put the probability of choosing one of the best compounds above $50\%$. Further splits are then necessary. 
  The optimal route is completed by ensuring that the element acting as $X$ has $\chi$ coming from the right-hand half of the periodic table and smaller than $1.1$.
  The  difference {in criteria} between the two studied temperatures is expected considering the more even mixture of $n$-type and $p$-type doped materials at $1000\,\mathrm{K}$, 
  and the increased relevance of electrical transport properties at that temperature. {As the contribution of electrons to thermal conduction gets higher, the role of the power factor in determining the best performing compounds increases. In view of the data, the condition $\Delta Z_{AB}<10$ is clearly connected with low power factors.}

To complete the analysis, we also examine the effect of having particular elements either in the $\left\lbrace A,B \right\rbrace$ positions or as $X$.
The most effective elements are those pushing the compound beyond the 75th percentile of the $ZT$ distribution.
We obtain the probabilities for positions' types and for temperatures. We select the elements
improving the probability above the average ($25\%$ by percentile definition). {We also impose the constraint that such elements must appear in at least two compounds in the correct position.}
At $300\,\mathrm{K}$ the elements most likely to yield a good thermoelectric HH are \{{V,Ga,Na,Al,Si,Zr,Sn}\} as $A$ or $B$, and {Co}
as $X$. At $1000\,\mathrm{K}$ they are \{Na,Al,Si{,Ti,Sn}\} and Co, respectively.

The full list of conditional probabilities is provided in the  {supporting} information. {The noticeable overlap between the best candidates at both temperatures is an{other} indicator of the correlation between room- and high-temperature behaviors. This is advantageous from an experimental point of view, as it means that some HHs may be tuned to yield good thermoelectric efficiencies at a wide range of temperatures.}

Overall, the best five compounds are:

{\bfseries Room-$T$:} {BiBaK, SbNaSr, AuAlHf, CoBiZr and RhSnTa.}

{\bfseries High-$T$:} {BiBaK, RhSnTa, AuAlHf, CoBiZr and CoAsHf}

In conclusion, we have extracted and examined the {$75$ thermodynamically} stable HH semiconductors out of a library of $79,057$ {\small \sf AFLOWLIB} HH entries.
We have characterized their thermoelectric properties in nanograined form using a completely {\itshape ab-initio} approach.

The results show that nanograined HHs have significantly higher overall $ZT$s than IV and III-V semiconductors.
A large number of potential compounds with {$ZT>0.5$} at room temperature and {$ZT>2$} at high temperatures
are competitive with the state of the art.
Such high values of $ZT$ are caused by average thermal conductivities and very high power factors {in the nanograined regime}.

The two key properties for high $ZT$ are a large lattice parameter and either a wide gap (at high temperatures), or a large effective mass of holes (at room temperature).
The presented recursive partitioning algorithm unveils {simple recipes to choose, with high probability, a good nanograined thermoelectric half-Heusler in either of the two temperature regimes}.
These practical selection criteria can stimulate experimental research for improving thermoelectric performance of HH semiconductors.

\section*{Methods}
{\bfseries AFLOWLIB repository of half-Heusler systems.}
The $79,057$ half-Heusler systems are calculated with the high-throughput framework \cite{aflowTHERMO,aflowPAPER,aflowBZ,aflowSCINT} {\small AFLOW}
based on {\itshape ab-initio} calculations of the energies by the {\scshape vasp} software \cite{kresse_vasp}. Projector augmented waves (PAW)
pseudopotentials \cite{PAW}, and Perdew, Burke and Ernzerhof exchange-correlation functionals \cite{PBE} are used.
The {\small \sf AFLOWLIB} energies are calculated at zero temperature and pressure, with spin polarization and without zero-point motion or lattice vibrations.
All crystal structures are fully relaxed (cell volume, spin, shape, and the basis atom coordinates inside the cell). 
Initial conditions were ferromagnetic; no antiferromagnetic configurations were tried because of the low magnetic ordering (if any) expected in semiconductors. 
Numerical convergence to about 1 meV/atom
is ensured by a high energy cutoff ($30\%$ higher than the highest energy cutoff for the pseudo-potentials of the components), and by the dense $6,000$
{\bfseries k}-points per reciprocal atom Monkhorst-Pack meshes \cite{monkhorst}. 
For each compound within the {\small \sf AFLOWLIB} repository we extract the symmetries, band structures, band gaps, and effective masses of electrons and holes.
The data can be freely downloaded from the consortium repository {\small \sf AFLOWLIB}.org directly or 
automatically by using the standardized application program interface \cite{curtarolo:art92}.

{\bfseries Transport coefficients.}
In the nanograined limit we have:
\begin{subequations}\label{eqn2}
  \begin{align}
    \sigma&=\lambda\frac{e^2}{k_{\mathrm{B}}T} M_0^{\left(\mathrm{FD}\right)},\label{eqn2a}\\
    S&=\frac{\lambda}{\sigma}\frac{e}{k_{\mathrm{B}}T^2} M_1^{\left(\mathrm{FD}\right)},\label{eqn2b}\\
    \kappa_{\mathrm{e}}&=\lambda\frac{1}{k_{\mathrm{B}}T^2}M_2^{\left(\mathrm{FD}\right)}-\sigma S^2T,\label{eqn2c}\\
    \kappa_{\ell}&=\lambda\frac{1}{k_{\mathrm{B}}T^2} M_2^{\left(\mathrm{BE}\right)}.\label{eqn2d}
  \end{align}
\end{subequations}
\noindent Here, $k_{\mathrm{B}}$ is the Boltzmann constant and $e$ the elementary unit of charge, and we define the integrals:
\begin{subequations}\label{eqn3}
  \begin{align}
    M_n^{\left(\mathrm{FD}\right)}&=\sum\limits_{\alpha}\int\limits_{BZ}f_{\mathrm{FD}}\left(f_{\mathrm{FD}}-1\right)\left\vert v^{\left(z\right)}_{\alpha,\mathbf{k}}\right\vert\left(\epsilon_{\alpha,\mathbf{k}}-\mu\right)^n\frac{d^3\vec{k}}{\left(2\pi\right)^3},\label{eqn3a}\\
    M_n^{\left(\mathrm{BE}\right)}&=\sum\limits_{\alpha}\int\limits_{BZ}f_{\mathrm{BE}}\left(f_{\mathrm{BE}}+1\right)\left\vert v^{\left(z\right)}_{\alpha,\mathbf{q}}\right\vert\left(\hbar\omega_{\alpha,\mathbf{q}}\right)^n\frac{d^3\vec{q}}{\left(2\pi\right)^3},\label{eqn3b}
  \end{align}
\end{subequations}
\noindent where $f_{\mathrm{FD}}$ ($f_{\mathrm{BE}}$) is the Fermi-Dirac (Bose-Einstein) distribution, the $\alpha$ index runs over electronic (phonon) bands,
$v^{\left(z\right)}_{\alpha,\mathbf{k}}$ ($v^{\left(z\right)}_{\alpha,\mathbf{q}}$) is the electronic (phonon) group velocity in the transport direction and
$\epsilon_{\alpha,\mathbf{k}}-\mu$ ($\hbar\omega_{\alpha,\mathbf{q}}$) is the difference between the energy of the corresponding carriers and the chemical potential, which is zero for phonons.

{\bfseries Interatomic force constants (IFCs) and phonon dispersions.}
$3\!\times\!3\!\times\!3$ supercells are used in second-order IFC calculations. The Phonopy \cite{phonopy} package is used to generate a minimal
set of atomic displacements by harnessing the space group of the crystal structure. $3\!\times\!3\!\times\!3$ Monkhorst-Pack {\bfseries k}-point
grids are employed. Spin polarization is excluded to improve speed. Phonon dispersions are obtained as the square roots of the eigenvalues of the
dynamical matrix. This matrix is built by combining the Fourier transform of the IFC matrix with a non-analytic correction to account for long-range Coulomb interactions
and reproduce LO-TO splitting \cite{Wang2010}. The ingredients for the latter are the dielectric tensor of the solid and a set of Born effective charges,
obtained using density functional perturbation theory as implemented in {\scshape vasp}.
We employ $32\!\times\!32\!\times\!32$ Monkhorst-Pack $\mathbf{q}$-point integration grids in the Brillouin zone to approximate the
integrals $M_2^{\left(\mathrm{BE}\right)}$ needed for Eq. (2d).

{\bfseries Electronic transport properties.}
To calculate $M_n^{\left(\mathrm{FD}\right)}$, defined by Eq. (3a), we have customized the {\scshape BoltzTraP} \cite{boltztrap} to work under the constant-mean-free-path ansatz.
The patched version of {\scshape BoltzTraP} is available from \url{https://bitbucket.org/sousaw/boltztrap-lambda}. {\scshape BoltzTraP} is based on a smoothed Fourier interpolation of the electronic bands.
As input, we provide electronic eigenenergies for each system computed on a $80\!\times\!80\!\times\!80$ Monkhorst-Pack $\mathbf{k}$-point grid.

{\bfseries Doping.} The effect of doping is simulated with rigid displacements of the chemical potential. For each compound and each temperature,
we select the doping level optimizing $ZT$. As mentioned before, the carrier concentration is $\le 10^{21}\,\mathrm{cm^{-3}}$,
within the doping limits typical of experimental HH thermoelectrics \cite{snyder_complex_2008}. {Numerous references have shown this approximation to be applicable in the typical regimes of thermoelectric interest.\cite{Restrepo2009,SelloniDoping,lee_tight-binding_2006,aflowTHERMO,aflowKAPPA}}

\begin{acknowledgement}
  {This work is partially supported by the French ``Carnot'' project SIEVE, by DOD-ONR (N00014-13-1-0635, N00014-11-1-0136, and N00014-09-1-0921), 
    by  DOE (DE-AC02-05CH11231, BES \#EDCBEE)  and by Duke University -- Center for Materials Genomics.} 
  The consortium {\small \sf AFLOWLIB}.org acknowledges the CRAY corporation for computational assistance.
\end{acknowledgement}

{\bf Supplementary information}
{Full nanograined-limit thermoelectric properties for all compounds as a function of temperature, estimates of nanograin sizes and conditional probabilities that the presence of an element in a compund pushes it beyond the 75th percentile of the $ZT$ distribution.}

\small
\newcommand{\Ozolins}{Ozoli\c{n}\v{s}}
\providecommand{\latin}[1]{#1}
\providecommand*\mcitethebibliography{\thebibliography}
\csname @ifundefined\endcsname{endmcitethebibliography}
  {\let\endmcitethebibliography\endthebibliography}{}


\begin{mcitethebibliography}{62}
\providecommand*\natexlab[1]{#1}
\providecommand*\mciteSetBstSublistMode[1]{}
\providecommand*\mciteSetBstMaxWidthForm[2]{}
\providecommand*\mciteBstWouldAddEndPuncttrue
  {\def\EndOfBibitem{\unskip.}}
\providecommand*\mciteBstWouldAddEndPunctfalse
  {\let\EndOfBibitem\relax}
\providecommand*\mciteSetBstMidEndSepPunct[3]{}
\providecommand*\mciteSetBstSublistLabelBeginEnd[3]{}
\providecommand*\EndOfBibitem{}
\mciteSetBstSublistMode{f}
\mciteSetBstMaxWidthForm{subitem}{(\alph{mcitesubitemcount})}
\mciteSetBstSublistLabelBeginEnd
  {\mcitemaxwidthsubitemform\space}
  {\relax}
  {\relax}

\bibitem[Snyder and Toberer(2008)Snyder, and Toberer]{snyder_complex_2008}
Snyder,~G.~J.; Toberer,~E.~S. Complex thermoelectric materials. \emph{Nat.\
  Mater.} \textbf{2008}, \emph{7}, 105--114\relax
\mciteBstWouldAddEndPuncttrue
\mciteSetBstMidEndSepPunct{\mcitedefaultmidpunct}
{\mcitedefaultendpunct}{\mcitedefaultseppunct}\relax
\EndOfBibitem
\bibitem[Curtarolo \latin{et~al.}(2013)Curtarolo, Hart, {Buongiorno~Nardelli},
  Mingo, Sanvito, and Levy]{curtarolo:nmat_review}
Curtarolo,~S.; Hart,~G.~L.~W.; {Buongiorno~Nardelli},~M.; Mingo,~N.;
  Sanvito,~S.; Levy,~O. The high-throughput highway to computational materials
  design. \emph{Nat.\ Mater.} \textbf{2013}, \emph{12}, 191--201\relax
\mciteBstWouldAddEndPuncttrue
\mciteSetBstMidEndSepPunct{\mcitedefaultmidpunct}
{\mcitedefaultendpunct}{\mcitedefaultseppunct}\relax
\EndOfBibitem
\bibitem[Snyder and Ursell(2003)Snyder, and Ursell]{snyder_thermoelectric_2003}
Snyder,~G.~J.; Ursell,~T.~S. Thermoelectric Efficiency and Compatibility.
  \emph{Phys.\ Rev.\ Lett.} \textbf{2003}, \emph{91}, 148301\relax
\mciteBstWouldAddEndPuncttrue
\mciteSetBstMidEndSepPunct{\mcitedefaultmidpunct}
{\mcitedefaultendpunct}{\mcitedefaultseppunct}\relax
\EndOfBibitem
\bibitem[Vining(2009)]{Vining_NM_2009}
Vining,~C.~B. An inconvenient truth about thermoelectrics. \emph{Nat.\ Mater.}
  \textbf{2009}, \emph{8}, 83--85\relax
\mciteBstWouldAddEndPuncttrue
\mciteSetBstMidEndSepPunct{\mcitedefaultmidpunct}
{\mcitedefaultendpunct}{\mcitedefaultseppunct}\relax
\EndOfBibitem
\bibitem[Li \latin{et~al.}(2010)Li, Liu, Zhao, and
  Zhou]{li_high-performance_2010}
Li,~J.-F.; Liu,~W.-S.; Zhao,~L.-D.; Zhou,~M. High-performance nanostructured
  thermoelectric materials. \emph{{NPG} Asia Mater.} \textbf{2010}, \emph{2},
  152--158\relax
\mciteBstWouldAddEndPuncttrue
\mciteSetBstMidEndSepPunct{\mcitedefaultmidpunct}
{\mcitedefaultendpunct}{\mcitedefaultseppunct}\relax
\EndOfBibitem
\bibitem[Curtin \latin{et~al.}(2012)Curtin, Fang, and
  Bowers]{curtin_highly_2012}
Curtin,~B.~M.; Fang,~E.~W.; Bowers,~J.~E. Highly Ordered Vertical Silicon
  Nanowire Array Composite Thin Films for Thermoelectric Devices. \emph{J.
  Electron. Mater.} \textbf{2012}, \emph{41}, 887--894\relax
\mciteBstWouldAddEndPuncttrue
\mciteSetBstMidEndSepPunct{\mcitedefaultmidpunct}
{\mcitedefaultendpunct}{\mcitedefaultseppunct}\relax
\EndOfBibitem
\bibitem[Pei \latin{et~al.}(2011)Pei, Lensch-Falk, Toberer, Medlin, and
  Snyder]{pei_high_2011}
Pei,~Y.; Lensch-Falk,~J.; Toberer,~E.~S.; Medlin,~D.~L.; Snyder,~G.~J. High
  Thermoelectric Performance in {PbTe} Due to Large Nanoscale {Ag$_2$Te}
  Precipitates and La Doping. \emph{Adv.\ Func.\ Mater.} \textbf{2011},
  \emph{21}, 241--249\relax
\mciteBstWouldAddEndPuncttrue
\mciteSetBstMidEndSepPunct{\mcitedefaultmidpunct}
{\mcitedefaultendpunct}{\mcitedefaultseppunct}\relax
\EndOfBibitem
\bibitem[Bux \latin{et~al.}(2009)Bux, Blair, Gogna, Lee, Chen, Dresselhaus,
  Kaner, and Fleurial]{bux_nanostructured_2009}
Bux,~S.~K.; Blair,~R.~G.; Gogna,~P.~K.; Lee,~H.; Chen,~G.; Dresselhaus,~M.~S.;
  Kaner,~R.~B.; Fleurial,~J.-P. Nanostructured Bulk Silicon as an Effective
  Thermoelectric Material. \emph{Adv.\ Func.\ Mater.} \textbf{2009}, \emph{19},
  2445--2452\relax
\mciteBstWouldAddEndPuncttrue
\mciteSetBstMidEndSepPunct{\mcitedefaultmidpunct}
{\mcitedefaultendpunct}{\mcitedefaultseppunct}\relax
\EndOfBibitem
\bibitem[Lan \latin{et~al.}(2009)Lan, Poudel, Ma, Wang, Dresselhaus, Chen, and
  Ren]{lan_structure_2009}
Lan,~Y.; Poudel,~B.; Ma,~Y.; Wang,~D.; Dresselhaus,~M.~S.; Chen,~G.; Ren,~Z.~F.
  Structure study of bulk nanograined thermoelectric bismuth antimony
  telluride. \emph{Nano\ Lett.} \textbf{2009}, \emph{9}, 1419--1422\relax
\mciteBstWouldAddEndPuncttrue
\mciteSetBstMidEndSepPunct{\mcitedefaultmidpunct}
{\mcitedefaultendpunct}{\mcitedefaultseppunct}\relax
\EndOfBibitem
\bibitem[Poon \latin{et~al.}(2013)Poon, Petersen, and
  Wu]{Poon_coreshell_APL_2013}
Poon,~S.~J.; Petersen,~A.~S.; Wu,~D. Thermal conductivity of core-shell
  nanocomposites for enhancing thermoelectric performance. \emph{Appl.\ Phys.\
  Lett.} \textbf{2013}, \emph{102}, 173110\relax
\mciteBstWouldAddEndPuncttrue
\mciteSetBstMidEndSepPunct{\mcitedefaultmidpunct}
{\mcitedefaultendpunct}{\mcitedefaultseppunct}\relax
\EndOfBibitem
\bibitem[Sootsman \latin{et~al.}(2009)Sootsman, He, Dravid, Li, Uher, and
  Kanatzidis]{Sootsman_JAP_2009}
Sootsman,~J.~R.; He,~J.; Dravid,~V.~P.; Li,~C.-P.; Uher,~C.; Kanatzidis,~M.~G.
  High thermoelectric figure of merit and improved mechanical properties in
  melt quenched PbTe-Ge and PbTe-Ge$_{1-x}$Si$_x$ eutectic and hypereutectic
  composites. \emph{J.\ Appl.\ Phys.} \textbf{2009}, \emph{105}, 083718\relax
\mciteBstWouldAddEndPuncttrue
\mciteSetBstMidEndSepPunct{\mcitedefaultmidpunct}
{\mcitedefaultendpunct}{\mcitedefaultseppunct}\relax
\EndOfBibitem
\bibitem[Gelbstein \latin{et~al.}(2010)Gelbstein, Dado, Ben-Yehuda, Sadia, Z,
  and Dariel]{Gelbstein_CHEMMAT_Spinodal_2010}
Gelbstein,~Y.; Dado,~B.; Ben-Yehuda,~O.; Sadia,~Y.; Z,~D.; Dariel,~M.~P. High
  Thermoelectric Figure of Merit and Nanostructuring in Bulk {\it p}-type
  Ge$_x$(Sn$_y$Pb$_{1-y}$)$_{1-x}$Te Alloys Following a Spinodal Decomposition
  Reaction. \emph{Chem.\ Mater.} \textbf{2010}, \emph{22}, 1054--1058\relax
\mciteBstWouldAddEndPuncttrue
\mciteSetBstMidEndSepPunct{\mcitedefaultmidpunct}
{\mcitedefaultendpunct}{\mcitedefaultseppunct}\relax
\EndOfBibitem
\bibitem[Poudel \latin{et~al.}(2008)Poudel, Hao, Ma, Lan, Minnich, Yu, Yan,
  Wang, Muto, Vashaee, Chen, Liu, Dresselhaus, Chen, and Ren]{Poudel2008}
Poudel,~B.; Hao,~Q.; Ma,~Y.; Lan,~Y.; Minnich,~A.; Yu,~B.; Yan,~X.; Wang,~D.;
  Muto,~A.; Vashaee,~D. \latin{et~al.}  High-thermoelectric performance of
  nanostructured bismuth antimony telluride bulk alloys. \emph{Science}
  \textbf{2008}, \emph{320}, 634--638\relax
\mciteBstWouldAddEndPuncttrue
\mciteSetBstMidEndSepPunct{\mcitedefaultmidpunct}
{\mcitedefaultendpunct}{\mcitedefaultseppunct}\relax
\EndOfBibitem
\bibitem[Biswas \latin{et~al.}(2012)Biswas, He, Blum, Wu, Hogan, Seidman,
  Dravid, and Kanatzidis]{biswas_high-performance_2012}
Biswas,~K.; He,~J.; Blum,~I.~D.; Wu,~C.-I.; Hogan,~T.~P.; Seidman,~D.~N.;
  Dravid,~V.~P.; Kanatzidis,~M.~G. High-performance bulk thermoelectrics with
  all-scale hierarchical architectures. \emph{Nature} \textbf{2012},
  \emph{489}, 414--418\relax
\mciteBstWouldAddEndPuncttrue
\mciteSetBstMidEndSepPunct{\mcitedefaultmidpunct}
{\mcitedefaultendpunct}{\mcitedefaultseppunct}\relax
\EndOfBibitem
\bibitem[Hochbaum \latin{et~al.}(2008)Hochbaum, Chen, Delgado, Liang, Garnett,
  Najarian, Majumdar, and Yang]{hochbaum_enhanced_2008}
Hochbaum,~A.~I.; Chen,~R.; Delgado,~R.~D.; Liang,~W.; Garnett,~E.~C.;
  Najarian,~M.; Majumdar,~A.; Yang,~P. Enhanced thermoelectric performance of
  rough silicon nanowires. \emph{Nature} \textbf{2008}, \emph{451},
  163--167\relax
\mciteBstWouldAddEndPuncttrue
\mciteSetBstMidEndSepPunct{\mcitedefaultmidpunct}
{\mcitedefaultendpunct}{\mcitedefaultseppunct}\relax
\EndOfBibitem
\bibitem[Madsen(2006)]{madsen2006}
Madsen,~G.~K.~H. Automated Search for New Thermoelectric Materials: The Case of
  LiZnSb. \emph{J.\ Am.\ Chem.\ Soc.} \textbf{2006}, \emph{128},
  12140--12146\relax
\mciteBstWouldAddEndPuncttrue
\mciteSetBstMidEndSepPunct{\mcitedefaultmidpunct}
{\mcitedefaultendpunct}{\mcitedefaultseppunct}\relax
\EndOfBibitem
\bibitem[Chepulskii and Curtarolo(2009)Chepulskii, and
  Curtarolo]{curtarolo:art41}
Chepulskii,~R.~V.; Curtarolo,~S. First-principles solubilities of alkali and
  alkaline-earth metals in Mg-B alloys. \emph{Phys.\ Rev.\ B} \textbf{2009},
  \emph{79}, 134203\relax
\mciteBstWouldAddEndPuncttrue
\mciteSetBstMidEndSepPunct{\mcitedefaultmidpunct}
{\mcitedefaultendpunct}{\mcitedefaultseppunct}\relax
\EndOfBibitem
\bibitem[Xie \latin{et~al.}(2012)Xie, Weidenkaff, Tang, Zhang, S.~J.~Poon, and
  Tritt]{xie_recent_2012}
Xie,~W.; Weidenkaff,~A.; Tang,~X.; Zhang,~Q.; S.~J.~Poon,~J.; Tritt,~T.~M.
  Recent Advances in Nanostructured Thermoelectric half-{Heusler} Compounds.
  \emph{Nanomaterials} \textbf{2012}, \emph{2}, 379--412\relax
\mciteBstWouldAddEndPuncttrue
\mciteSetBstMidEndSepPunct{\mcitedefaultmidpunct}
{\mcitedefaultendpunct}{\mcitedefaultseppunct}\relax
\EndOfBibitem
\bibitem[Yan \latin{et~al.}(2011)Yan, Joshi, Liu, Lan, Wang, Lee, Simonson,
  Poon, Tritt, Chen, and Ren]{Yan_NL_2011}
Yan,~X.; Joshi,~G.; Liu,~W.; Lan,~Y.; Wang,~H.; Lee,~S.; Simonson,~J.~W.;
  Poon,~S.~J.; Tritt,~T.~M.; Chen,~G. \latin{et~al.}  Enhanced Thermoelectric
  figure of merit of p-Type half-{Heuslers}. \emph{Nano\ Lett.} \textbf{2011},
  \emph{11}, 556--560\relax
\mciteBstWouldAddEndPuncttrue
\mciteSetBstMidEndSepPunct{\mcitedefaultmidpunct}
{\mcitedefaultendpunct}{\mcitedefaultseppunct}\relax
\EndOfBibitem
\bibitem[Poon \latin{et~al.}(2011)Poon, Wu, Zhu, Xie, Tritt, Thomas, and
  Venkatasubramanian]{poon_half-heusler_2011}
Poon,~S.~J.; Wu,~D.; Zhu,~S.; Xie,~W.; Tritt,~T.~M.; Thomas,~P.;
  Venkatasubramanian,~R. {H}alf-{H}eusler phases and nanocomposites as emerging
  high-{ZT} thermoelectric materials. \textbf{2011}, \emph{26},
  2795--2802\relax
\mciteBstWouldAddEndPuncttrue
\mciteSetBstMidEndSepPunct{\mcitedefaultmidpunct}
{\mcitedefaultendpunct}{\mcitedefaultseppunct}\relax
\EndOfBibitem
\bibitem[Uher \latin{et~al.}(1999)Uher, Yang, and
  Meisner]{uher_thermoelectric_1999}
Uher,~C.; Yang,~J.; Meisner,~G.~P. Thermoelectric properties of Bi-doped
  half-{Heusler} alloys. Eighteenth International Conference on
  Thermoelectrics, 1999, doi=10.1109/ICT.1999.843333. 1999; pp 56--59\relax
\mciteBstWouldAddEndPuncttrue
\mciteSetBstMidEndSepPunct{\mcitedefaultmidpunct}
{\mcitedefaultendpunct}{\mcitedefaultseppunct}\relax
\EndOfBibitem
\bibitem[Culp \latin{et~al.}(2008)Culp, Simonson, Poon, Ponnambalam, Edwards,
  and Tritt]{culp_zrhfcosbsn_2008}
Culp,~S.~R.; Simonson,~J.~W.; Poon,~S.~J.; Ponnambalam,~V.; Edwards,~J.;
  Tritt,~T.~M. ({Zr,Hf)Co(Sb}{,Sn)} half-{Heusler} phases as high-temperature
  (>700$^{\circ}$C) p-type thermoelectric materials. \emph{Appl.\ Phys.\ Lett.}
  \textbf{2008}, \emph{93}, 022105\relax
\mciteBstWouldAddEndPuncttrue
\mciteSetBstMidEndSepPunct{\mcitedefaultmidpunct}
{\mcitedefaultendpunct}{\mcitedefaultseppunct}\relax
\EndOfBibitem
\bibitem[Yang \latin{et~al.}(2008)Yang, Li, Wu, Zhang, Chen, and
  Yang]{yang_evaluation_2008}
Yang,~J.; Li,~H.; Wu,~T.; Zhang,~W.; Chen,~L.; Yang,~J. Evaluation of
  half-{Heusler} Compounds as Thermoelectric Materials Based on the Calculated
  Electrical Transport Properties. \emph{Adv.\ Func.\ Mater.} \textbf{2008},
  \emph{18}, 2880--2888\relax
\mciteBstWouldAddEndPuncttrue
\mciteSetBstMidEndSepPunct{\mcitedefaultmidpunct}
{\mcitedefaultendpunct}{\mcitedefaultseppunct}\relax
\EndOfBibitem
\bibitem[Wee \latin{et~al.}(2012)Wee, Kozinsky, Pavan, and
  Fornari]{Wee_Fornari_TiNiSn_JEM_2012}
Wee,~D.; Kozinsky,~B.; Pavan,~B.; Fornari,~M. Quasiharmonic Vibrational
  Properties of {TiNiSn} from {\it Ab-Initio} Phonons. \emph{J.\ Elec.\ Mat.}
  \textbf{2012}, \emph{41}, 977--983\relax
\mciteBstWouldAddEndPuncttrue
\mciteSetBstMidEndSepPunct{\mcitedefaultmidpunct}
{\mcitedefaultendpunct}{\mcitedefaultseppunct}\relax
\EndOfBibitem
\bibitem[Matsumoto \latin{et~al.}(2003)Matsumoto, Koinuma, Hasegawa, Takeuchi,
  Tsui, and Yoo]{matsumoto2003combinatorial}
Matsumoto,~Y.; Koinuma,~H.; Hasegawa,~T.; Takeuchi,~I.; Tsui,~F.; Yoo,~Y.~K.
  Combinatorial investigation of spintronic materials. \emph{MRS bulletin}
  \textbf{2003}, \emph{28}, 734--739\relax
\mciteBstWouldAddEndPuncttrue
\mciteSetBstMidEndSepPunct{\mcitedefaultmidpunct}
{\mcitedefaultendpunct}{\mcitedefaultseppunct}\relax
\EndOfBibitem
\bibitem[Yang \latin{et~al.}(2012)Yang, Setyawan, Wang, {Buongiorno~Nardelli},
  and Curtarolo]{curtarolo:TIs}
Yang,~K.; Setyawan,~W.; Wang,~S.; {Buongiorno~Nardelli},~M.; Curtarolo,~S. A
  search model for topological insulators with high-throughput robustness
  descriptors. \emph{Nat.\ Mater.} \textbf{2012}, \emph{11}, 614--619\relax
\mciteBstWouldAddEndPuncttrue
\mciteSetBstMidEndSepPunct{\mcitedefaultmidpunct}
{\mcitedefaultendpunct}{\mcitedefaultseppunct}\relax
\EndOfBibitem
\bibitem[Curtarolo \latin{et~al.}(2012)Curtarolo, Setyawan, Wang, Xue, Yang,
  Taylor, Nelson, Hart, Sanvito, {Buongiorno~Nardelli}, Mingo, and
  Levy]{aflowlibPAPER}
Curtarolo,~S.; Setyawan,~W.; Wang,~S.; Xue,~J.; Yang,~K.; Taylor,~R.~H.;
  Nelson,~L.~J.; Hart,~G.~L.~W.; Sanvito,~S.; {Buongiorno~Nardelli},~M.
  \latin{et~al.}  AFLOWLIB.ORG: A distributed materials properties repository
  from high-throughput {\it ab initio} calculations. \emph{Comp.\ Mat.\ Sci.}
  \textbf{2012}, \emph{58}, 227--235\relax
\mciteBstWouldAddEndPuncttrue
\mciteSetBstMidEndSepPunct{\mcitedefaultmidpunct}
{\mcitedefaultendpunct}{\mcitedefaultseppunct}\relax
\EndOfBibitem
\bibitem[{S.~Curtarolo, ~W.~Setyawan, ~G.~L.~W.~Hart, ~M.~Jahnatek,
  ~R.~V.~Chepulskii, ~R.~H.~Taylor, ~S.~Wang, ~J.~Xue, ~K.~Yang, ~O.~Levy,
  ~M.~Mehl, ~H.~T.~Stokes, ~D.~O.~Demchenko, and~D.~Morgan}(2012)]{aflowPAPER}
{S.~Curtarolo, ~W.~Setyawan, ~G.~L.~W.~Hart, ~M.~Jahnatek, ~R.~V.~Chepulskii,
  ~R.~H.~Taylor, ~S.~Wang, ~J.~Xue, ~K.~Yang, ~O.~Levy, ~M.~Mehl,
  ~H.~T.~Stokes, ~D.~O.~Demchenko, and~D.~Morgan}, AFLOW: an automatic
  framework for high-throughput materials discovery. \emph{Comp.\ Mat.\ Sci.}
  \textbf{2012}, \emph{58}, 218--226\relax
\mciteBstWouldAddEndPuncttrue
\mciteSetBstMidEndSepPunct{\mcitedefaultmidpunct}
{\mcitedefaultendpunct}{\mcitedefaultseppunct}\relax
\EndOfBibitem
\bibitem[Bergerhoff \latin{et~al.}(1983)Bergerhoff, Hundt, Sievers, and
  Brown]{ICSD}
Bergerhoff,~G.; Hundt,~R.; Sievers,~R.; Brown,~I.~D. The inorganic crystal
  structure data base. \emph{J. Chem. Inf. Comput. Sci.} \textbf{1983},
  \emph{23}, 66--69\relax
\mciteBstWouldAddEndPuncttrue
\mciteSetBstMidEndSepPunct{\mcitedefaultmidpunct}
{\mcitedefaultendpunct}{\mcitedefaultseppunct}\relax
\EndOfBibitem
\bibitem[{FIZ~Karlsruhe~and~NIST}()]{ICSD4}
{FIZ~Karlsruhe~and~NIST}, Inorganic Crystal Structure Database.
  http://icsd.fiz-karlsruhe.de/icsd/\relax
\mciteBstWouldAddEndPuncttrue
\mciteSetBstMidEndSepPunct{\mcitedefaultmidpunct}
{\mcitedefaultendpunct}{\mcitedefaultseppunct}\relax
\EndOfBibitem
\bibitem[Bazela and Szytula(1981)Bazela, and Szytula]{NiMnTiGe_Bazela_1981}
Bazela,~W.; Szytula,~A. Crystal and magnetic structure of the
  {NiMn$_{1−t}$Ti$_t$Ge} system. \emph{Phys.\ Stat.\ Solidi\ A}
  \textbf{1981}, \emph{66}, 45--52\relax
\mciteBstWouldAddEndPuncttrue
\mciteSetBstMidEndSepPunct{\mcitedefaultmidpunct}
{\mcitedefaultendpunct}{\mcitedefaultseppunct}\relax
\EndOfBibitem
\bibitem[Carrete \latin{et~al.}(2014)Carrete, Li, Mingo, Wang, and
  Curtarolo]{curtarolo:art84}
Carrete,~J.; Li,~W.; Mingo,~N.; Wang,~S.; Curtarolo,~S. Finding unprecedentedly
  low-thermal-conductivity {half-Heusler} semiconductors via high-throughput
  materials modeling. \emph{Phys.\ Rev.\ X} \textbf{2014}, \emph{4},
  011019\relax
\mciteBstWouldAddEndPuncttrue
\mciteSetBstMidEndSepPunct{\mcitedefaultmidpunct}
{\mcitedefaultendpunct}{\mcitedefaultseppunct}\relax
\EndOfBibitem
\bibitem[Setyawan and Curtarolo(2010)Setyawan, and Curtarolo]{aflowBZ}
Setyawan,~W.; Curtarolo,~S. High-throughput electronic band structure
  calculations: challenges and tools. \emph{Comp.\ Mat.\ Sci.} \textbf{2010},
  \emph{49}, 299--312\relax
\mciteBstWouldAddEndPuncttrue
\mciteSetBstMidEndSepPunct{\mcitedefaultmidpunct}
{\mcitedefaultendpunct}{\mcitedefaultseppunct}\relax
\EndOfBibitem
\bibitem[Klemens(1994)]{klemens_phonon_1994}
Klemens,~P.~G. Phonon scattering and thermal resistance due to grain
  boundaries. \emph{Int. J. Thermophys.} \textbf{1994}, \emph{15},
  1345--1351\relax
\mciteBstWouldAddEndPuncttrue
\mciteSetBstMidEndSepPunct{\mcitedefaultmidpunct}
{\mcitedefaultendpunct}{\mcitedefaultseppunct}\relax
\EndOfBibitem
\bibitem[Callaway(1959)]{callaway_model_1959}
Callaway,~J. Model for Lattice Thermal Conductivity at Low Temperatures.
  \emph{Phys. Rev.} \textbf{1959}, \emph{113}, 1046--1051\relax
\mciteBstWouldAddEndPuncttrue
\mciteSetBstMidEndSepPunct{\mcitedefaultmidpunct}
{\mcitedefaultendpunct}{\mcitedefaultseppunct}\relax
\EndOfBibitem
\bibitem[Wang \latin{et~al.}(2011)Wang, Wang, Setyawan, Mingo, and
  Curtarolo]{aflowTHERMO}
Wang,~S.; Wang,~Z.; Setyawan,~W.; Mingo,~N.; Curtarolo,~S. Assessing the
  thermoelectric properties of sintered compounds via high-throughput ab-initio
  calculations. \emph{Phys.\ Rev.\ X} \textbf{2011}, \emph{1}, 021012\relax
\mciteBstWouldAddEndPuncttrue
\mciteSetBstMidEndSepPunct{\mcitedefaultmidpunct}
{\mcitedefaultendpunct}{\mcitedefaultseppunct}\relax
\EndOfBibitem
\bibitem[Mingo \latin{et~al.}(2014)Mingo, Stewart, Broido, Lindsay, and
  Li]{mingo_ab_2014}
Mingo,~N.; Stewart,~D.~A.; Broido,~D.~A.; Lindsay,~L.; Li,~W.
  \emph{Length-Scale Dependent Phonon Interactions}; Springer New York, 2014;
  pp 137--173\relax
\mciteBstWouldAddEndPuncttrue
\mciteSetBstMidEndSepPunct{\mcitedefaultmidpunct}
{\mcitedefaultendpunct}{\mcitedefaultseppunct}\relax
\EndOfBibitem
\bibitem[Tamura(1983)]{tamura_isotope_1983}
Tamura,~S. Isotope scattering of dispersive phonons in {Ge}. \emph{Phys.\ Rev.\
  B} \textbf{1983}, \emph{27}, 858--866\relax
\mciteBstWouldAddEndPuncttrue
\mciteSetBstMidEndSepPunct{\mcitedefaultmidpunct}
{\mcitedefaultendpunct}{\mcitedefaultseppunct}\relax
\EndOfBibitem
\bibitem[Kundu \latin{et~al.}(2011)Kundu, Mingo, Broido, and
  Stewart]{kundu_role_2011}
Kundu,~A.; Mingo,~N.; Broido,~D.~A.; Stewart,~D.~A. Role of light and heavy
  embedded nanoparticles on the thermal conductivity of {SiGe} alloys.
  \emph{Phys.\ Rev.\ B} \textbf{2011}, \emph{84}, 125426\relax
\mciteBstWouldAddEndPuncttrue
\mciteSetBstMidEndSepPunct{\mcitedefaultmidpunct}
{\mcitedefaultendpunct}{\mcitedefaultseppunct}\relax
\EndOfBibitem
\bibitem[Ziman(2001)]{ziman_electrons_2001}
Ziman,~J.~M. \emph{Electrons and Phonons: The Theory of Transport Phenomena in
  Solids}; Oxford University Press, 2001\relax
\mciteBstWouldAddEndPuncttrue
\mciteSetBstMidEndSepPunct{\mcitedefaultmidpunct}
{\mcitedefaultendpunct}{\mcitedefaultseppunct}\relax
\EndOfBibitem
\bibitem[Carrete \latin{et~al.}(2014)Carrete, Li, Mingo, Wang, and
  Curtarolo]{aflowKAPPA}
Carrete,~J.; Li,~W.; Mingo,~N.; Wang,~S.; Curtarolo,~S. Finding unprecedentedly
  low-thermal-conductivity {half-Heusler} semiconductors via high-throughput
  materials modeling. \emph{Phys.\ Rev.\ X} \textbf{2014}, \emph{4},
  011019\relax
\mciteBstWouldAddEndPuncttrue
\mciteSetBstMidEndSepPunct{\mcitedefaultmidpunct}
{\mcitedefaultendpunct}{\mcitedefaultseppunct}\relax
\EndOfBibitem
\bibitem[Wang \latin{et~al.}(2011)Wang, Alaniz, Jang, Garay, and
  Dames]{wang_thermal_2011}
Wang,~Z.; Alaniz,~J.~E.; Jang,~W.; Garay,~J.~E.; Dames,~C. Thermal Conductivity
  of Nanocrystalline Silicon: Importance of Grain Size and Frequency-Dependent
  Mean Free Paths. \emph{Nano\ Lett.} \textbf{2011}, \emph{11},
  2206--2213\relax
\mciteBstWouldAddEndPuncttrue
\mciteSetBstMidEndSepPunct{\mcitedefaultmidpunct}
{\mcitedefaultendpunct}{\mcitedefaultseppunct}\relax
\EndOfBibitem
\bibitem[Bachmann \latin{et~al.}(2012)Bachmann, Czerner, and
  Heiliger]{bachmann_ineffectiveness_2012}
Bachmann,~M.; Czerner,~M.; Heiliger,~C. Ineffectiveness of energy filtering at
  grain boundaries for thermoelectric materials. \emph{Phys.\ Rev.\ B}
  \textbf{2012}, \emph{86}, 115320\relax
\mciteBstWouldAddEndPuncttrue
\mciteSetBstMidEndSepPunct{\mcitedefaultmidpunct}
{\mcitedefaultendpunct}{\mcitedefaultseppunct}\relax
\EndOfBibitem
\bibitem[Parrott(1969)]{parrott_thermal_1969}
Parrott,~J.~E. The thermal conductivity of sintered semiconductor alloys.
  \emph{J.\ Phys.\ Chem.} \textbf{1969}, \emph{2}, 147\relax
\mciteBstWouldAddEndPuncttrue
\mciteSetBstMidEndSepPunct{\mcitedefaultmidpunct}
{\mcitedefaultendpunct}{\mcitedefaultseppunct}\relax
\EndOfBibitem
\bibitem[Pauling(1960)]{pauling_bond}
Pauling,~L. \emph{The Nature of the Chemical Bond and the Structure of
  Molecules and Crystals: An Introduction to Modern Structural Chemistry}, 3rd
  ed.; Cornell University Press: New York, 1960\relax
\mciteBstWouldAddEndPuncttrue
\mciteSetBstMidEndSepPunct{\mcitedefaultmidpunct}
{\mcitedefaultendpunct}{\mcitedefaultseppunct}\relax
\EndOfBibitem
\bibitem[Pettifor(1984)]{pettifor:1984}
Pettifor,~D.~G. A chemical scale for crystal-structure maps. \emph{Sol. State
  Commun.} \textbf{1984}, \emph{51}, 31--34\relax
\mciteBstWouldAddEndPuncttrue
\mciteSetBstMidEndSepPunct{\mcitedefaultmidpunct}
{\mcitedefaultendpunct}{\mcitedefaultseppunct}\relax
\EndOfBibitem
\bibitem[Breiman \latin{et~al.}(1984)Breiman, Friedman, Stone, and
  Olshen]{breiman_classification_1984}
Breiman,~L.; Friedman,~J.; Stone,~C.~J.; Olshen,~R.~A. \emph{Classification and
  Regression Trees}, 1st ed.; Chapman and {Hall/CRC}, 1984\relax
\mciteBstWouldAddEndPuncttrue
\mciteSetBstMidEndSepPunct{\mcitedefaultmidpunct}
{\mcitedefaultendpunct}{\mcitedefaultseppunct}\relax
\EndOfBibitem
\bibitem[Slack(1995)]{slack_crc_1995}
Slack,~G.~A. In \emph{{CRC} Handbook of Thermoelectrics}; Rowe,~D.~M., Ed.;
  {CRC} Press, 1995; Chapter New Materials and Performance Limits for
  Thermoelectric Cooling\relax
\mciteBstWouldAddEndPuncttrue
\mciteSetBstMidEndSepPunct{\mcitedefaultmidpunct}
{\mcitedefaultendpunct}{\mcitedefaultseppunct}\relax
\EndOfBibitem
\bibitem[Mahan(1997)]{Mahan_review_TE}
Mahan,~G.~D. In \emph{Good Thermoelectrics}; Ehrenreich,~H., Spaepen,~F., Eds.;
  Solid State Physics; Academic Press, 1997; Vol.~51; p~81\relax
\mciteBstWouldAddEndPuncttrue
\mciteSetBstMidEndSepPunct{\mcitedefaultmidpunct}
{\mcitedefaultendpunct}{\mcitedefaultseppunct}\relax
\EndOfBibitem
\bibitem[Setyawan \latin{et~al.}(2011)Setyawan, Gaume, Lam, Feigelson, and
  Curtarolo]{aflowSCINT}
Setyawan,~W.; Gaume,~R.~M.; Lam,~S.; Feigelson,~R.~S.; Curtarolo,~S.
  High-Throughput Combinatorial Database of Electronic Band Structures for
  Inorganic Scintillator Materials. \emph{ACS\ Comb.\ Sci.} \textbf{2011},
  \emph{13}, 382--390\relax
\mciteBstWouldAddEndPuncttrue
\mciteSetBstMidEndSepPunct{\mcitedefaultmidpunct}
{\mcitedefaultendpunct}{\mcitedefaultseppunct}\relax
\EndOfBibitem
\bibitem[Kresse and Hafner(1993)Kresse, and Hafner]{kresse_vasp}
Kresse,~G.; Hafner,~J. {\it Ab initio} molecular dynamics for liquid metals.
  \emph{Phys.\ Rev.\ B} \textbf{1993}, \emph{47}, 558--561\relax
\mciteBstWouldAddEndPuncttrue
\mciteSetBstMidEndSepPunct{\mcitedefaultmidpunct}
{\mcitedefaultendpunct}{\mcitedefaultseppunct}\relax
\EndOfBibitem
\bibitem[Bl\"ochl(1994)]{PAW}
Bl\"ochl,~P.~E. Projector augmented-wave method. \emph{Phys.\ Rev.\ B}
  \textbf{1994}, \emph{50}, 17953--17979\relax
\mciteBstWouldAddEndPuncttrue
\mciteSetBstMidEndSepPunct{\mcitedefaultmidpunct}
{\mcitedefaultendpunct}{\mcitedefaultseppunct}\relax
\EndOfBibitem
\bibitem[Perdew \latin{et~al.}(1996)Perdew, Burke, and Ernzerhof]{PBE}
Perdew,~J.~P.; Burke,~K.; Ernzerhof,~M. Generalized gradient approximation made
  simple. \emph{Phys.\ Rev.\ Lett.} \textbf{1996}, \emph{77}, 3865--3868\relax
\mciteBstWouldAddEndPuncttrue
\mciteSetBstMidEndSepPunct{\mcitedefaultmidpunct}
{\mcitedefaultendpunct}{\mcitedefaultseppunct}\relax
\EndOfBibitem
\bibitem[Monkhorst and Pack(1976)Monkhorst, and Pack]{monkhorst}
Monkhorst,~H.~J.; Pack,~J.~D. Special points for {Brillouin-zone} integrations.
  \emph{Phys.\ Rev.\ B} \textbf{1976}, \emph{13}, 5188--5192\relax
\mciteBstWouldAddEndPuncttrue
\mciteSetBstMidEndSepPunct{\mcitedefaultmidpunct}
{\mcitedefaultendpunct}{\mcitedefaultseppunct}\relax
\EndOfBibitem
\bibitem[Taylor \latin{et~al.}(2014)Taylor, Rose, Toher, Levy, Yang,
  {Buongiorno~Nardelli}, and Curtarolo]{curtarolo:art92}
Taylor,~R.~H.; Rose,~F.; Toher,~C.; Levy,~O.; Yang,~K.;
  {Buongiorno~Nardelli},~M.; Curtarolo,~S. A RESTful API for exchanging
  Materials Data in the AFLOWLIB.org consortium. \emph{Comp. Mat. Sci.
  doi=10.1016/j.commatsci.2014.05.014} \textbf{2014}, \relax
\mciteBstWouldAddEndPunctfalse
\mciteSetBstMidEndSepPunct{\mcitedefaultmidpunct}
{}{\mcitedefaultseppunct}\relax
\EndOfBibitem
\bibitem[Togo \latin{et~al.}(2008)Togo, Oba, and Tanaka]{phonopy}
Togo,~A.; Oba,~F.; Tanaka,~I. First-principles calculations of the ferroelastic
  transition between rutile-type and {CaCl$_2$}-type {SiO$_2$} at high
  pressures. \emph{Phys.\ Rev.\ B} \textbf{2008}, \emph{78}, 134106\relax
\mciteBstWouldAddEndPuncttrue
\mciteSetBstMidEndSepPunct{\mcitedefaultmidpunct}
{\mcitedefaultendpunct}{\mcitedefaultseppunct}\relax
\EndOfBibitem
\bibitem[Wang \latin{et~al.}(2010)Wang, Wang, Wang, Mei, Shang, Chen, and
  Liu]{Wang2010}
Wang,~Y.; Wang,~J.~J.; Wang,~W.~Y.; Mei,~Z.~G.; Shang,~S.~L.; Chen,~L.~Q.;
  Liu,~Z.~K. A mixed-space approach to first-principles calculations of phonon
  frequencies for polar materials. \emph{J. Phys.: Condens. Matter}
  \textbf{2010}, \emph{22}, 202201\relax
\mciteBstWouldAddEndPuncttrue
\mciteSetBstMidEndSepPunct{\mcitedefaultmidpunct}
{\mcitedefaultendpunct}{\mcitedefaultseppunct}\relax
\EndOfBibitem
\bibitem[Madsen and Singh(2006)Madsen, and Singh]{boltztrap}
Madsen,~G.~K.; Singh,~D.~J. {BoltzTraP}. A code for calculating band-structure
  dependent quantities. \emph{Computer Phys. Commun.} \textbf{2006},
  \emph{175}, 67--71\relax
\mciteBstWouldAddEndPuncttrue
\mciteSetBstMidEndSepPunct{\mcitedefaultmidpunct}
{\mcitedefaultendpunct}{\mcitedefaultseppunct}\relax
\EndOfBibitem
\bibitem[Restrepo \latin{et~al.}(2009)Restrepo, Varga, and
  Pantelides]{Restrepo2009}
Restrepo,~O.~D.; Varga,~K.; Pantelides,~S.~T. First-principles calculations of
  electron mobilities in silicon: Phonon and Coulomb scattering. \emph{Appl.\
  Phys.\ Lett.} \textbf{2009}, \emph{94}, 212103\relax
\mciteBstWouldAddEndPuncttrue
\mciteSetBstMidEndSepPunct{\mcitedefaultmidpunct}
{\mcitedefaultendpunct}{\mcitedefaultseppunct}\relax
\EndOfBibitem
\bibitem[Selloni and Pantelides(1982)Selloni, and Pantelides]{SelloniDoping}
Selloni,~A.; Pantelides,~S.~T. Electronic Structure and Spectra of Heavily
  Doped n-type Silicon. \emph{Phys.\ Rev.\ Lett.} \textbf{1982}, \emph{49},
  586\relax
\mciteBstWouldAddEndPuncttrue
\mciteSetBstMidEndSepPunct{\mcitedefaultmidpunct}
{\mcitedefaultendpunct}{\mcitedefaultseppunct}\relax
\EndOfBibitem
\bibitem[Lee and von Allmen(2006)Lee, and von Allmen]{lee_tight-binding_2006}
Lee,~S.; von Allmen,~P. Tight-binding modeling of thermoelectric properties of
  bismuth telluride. \emph{Appl.\ Phys.\ Lett.} \textbf{2006}, \emph{88},
  022107--3\relax
\mciteBstWouldAddEndPuncttrue
\mciteSetBstMidEndSepPunct{\mcitedefaultmidpunct}
{\mcitedefaultendpunct}{\mcitedefaultseppunct}\relax
\EndOfBibitem
\end{mcitethebibliography}
\end{document}